\documentclass[10pt,a4paper]{book}
\usepackage{udineHyderabad-Klu2}

\begin{document}

	\pagestyle{fancy}
	
\talktitle{Neural networks for gamma-hadron separation in MAGIC}{Neural networks for gamma-hadron separation in MAGIC}

\talkauthors{%
P.~Boinee\structure{a,b},
F.~Barbarino\structure{a,b},
A.~De Angelis\structure{a,b},
A.~Saggion\structure{c},
M.~Zacchello\structure{c}}

\authorstucture[a]{Dipartimento di Fisica,
                   Universit\`a di Udine,
                   via delle Scien\-ze~208, 33100~Udine, Italy}

\authorstucture[b]{INFN, Sezione di Trieste,
                   Gruppo di Udine,
                   via delle Scien\-ze~208, 33100~Udine, Italy}

\authorstucture[c]{Dipartimento di Fisica,
                   Universit\`a di Padova,
                   via Marzolo~8, 35131~Padova, Italy}

\shorttitle{Neural networks for gamma-hadron separation in MAGIC}

\firstauthor{P.~Boinee et al.}

		\index{Boinee@\textsc{Boinee}, P.}
		\index{Barbarino@\textsc{Barbarino}, F.}
		\index{De Angelis@\textsc{De Angelis}, A.}
		\index{Saggion@\textsc{Saggion}, A.}
		\index{Zacchello@\textsc{Zacchello}, M.}

\begin{abstract}
Neural networks have proved to be versatile and robust  for particle separation in many experiments related to particle astrophysics. We apply these techniques to separate  gamma rays from hadrons for the MAGIC \v Cerenkov Telescope. Two types of neural network architectures have been used for the classification task: one is the MultiLayer Perceptron (MLP) based on supervised learning, and the other is the Self-Organising Tree Algorithm (SOTA), which is based on unsupervised learning. We propose a new architecture by combining these two neural networks types to yield better and faster classification results for our classification problem.
\end{abstract}

\section{Introduction}
Many gamma ray experiments have to deal with the problem of separating gammas from hadrons. The experiments usually generate large data sets with many attributes in them. This multi-dimensional data classification problem offers a daunting challenge of extracting small number of interesting events (gammas) from an overwhelming sea of background (hadrons). Many techniques are in active research for addressing this problem. The list includes classical statistical multivariate techniques to more sophisticated techniques like neural networks, classification trees and kernel functions.

The class of neural networks provides an automated technique for the classification of the data set into given number of classes~\protect\cite{kk}. It is in active research in both artificial intelligence and machine learning communities.  Several neural network models have been developed to address the classification problem. Usually, one makes the distinction between supervised and unsupervised classifiers: the former are trained with data for which the classification is known and then used to classify raw data, while the latter attempt to find the best-fitting class structure in the input data by using some measure of merit (usually an euclidean metric is used~\protect\cite{raj}).
From a mathematical perspective, a neural network is simply a mapping from $R^n \rightarrow R^m$, where $R^n$ is the input data set dimension and $R^m$ is the output dimension of the neural network. The network is typically divided into various layers; each layer has a set of neurons also called nodes or information units, connected together by the links. The artificial neural networks are able to classify data by learning how to discriminate patterns in features (or parameters) associated with the data.
The neural network learns from the data set when each data vector from the input set is subjected to it. The learning or information gain is stored in the links associated with the neurons.
The output structure of the network is dependent on both the problem and the network type. For a gamma/hadron separation problem the network maps each input vector onto the [0,1] interval in supervised networks, whereas in unsupervised networks the nodes are adapted to the input vector in such a way that the output of the network represents the natural groups that exist in the data set. The output of the unsupervised network is generally stored in an ASCII file. A visualization technique is then used to view the groups by processing the output file generated by the network.

Section 2 describes the data sets used for the classification. Section 3 deals with the MultiLayer Perceptron network and its classification results. Section 4 deals with the Self-Organizing Tree Algorithm and its variant along with their classification results. Conclusions and future perspectives are discussed in the section 5.

\section{Data set description}
The data sets are generated by a MonteCarlo simulation program, CORSIKA~\protect\cite{cor}.
They contain 12332 gammas, 7356 ON events (mixture of gammas and hadrons), and 6688 hadron or OFF events. These events are stored in different files. The files contain event parameters in ASCII format, each line of 10 numbers being one event~\protect\cite{boc} with the parameters defined below.
\begin{enumerate}
\item {\em Length}:    major axis of ellipse [mm]
\item {\em Width}:     minor axis of ellipse [mm]
\item {\em Size}:      10-log of sum of content of all pixels
\item {\em Conc}:      ratio of sum of two highest pixels over fSize  [ratio]
\item {\em Conc1}:     ratio of highest pixel over fSize  [ratio]
\item {\em Asym}:      distance from highest pixel to centre, projected onto major axis [mm]
\item {\em M3Long}:    3rd root of third moment along major axis  [mm]
\item {\em M3Trans}:   3rd root of third moment along minor axis  [mm]
\item {\em Alpha}:     angle of major axis with vector to origin [deg]
\item {\em Dist}:      distance from origin to centre of ellipse [mm]
\end{enumerate}
These Hillas image parameters \protect\cite{hil} are derived from pixel analysis and are used for classification.

\section{Multi-Layer Perceptron}
For this approach we used the ROOT Analysis Package (v4.00/02) and in particular the MultiLayer Perceptron class~\protect\cite{mlp} which implements a generic layered network. Since this is a supervised network we took two thirds of gamma and OFF data to train the network and the remaining data to test it.The code of the ROOT package is very flexible and simple to use. It allowed us to create a network with a 10 nodes input layer, a hidden layer with the same number of nodes and an output layer with just a single neuron which should return "0" if the data represent hadrons or "1" if they are gammas. 
\begin{figure}[hb]
  \centering
  {\subfigure[The error functions for training and test data took on 1000 runs]{\includegraphics[width=6cm]{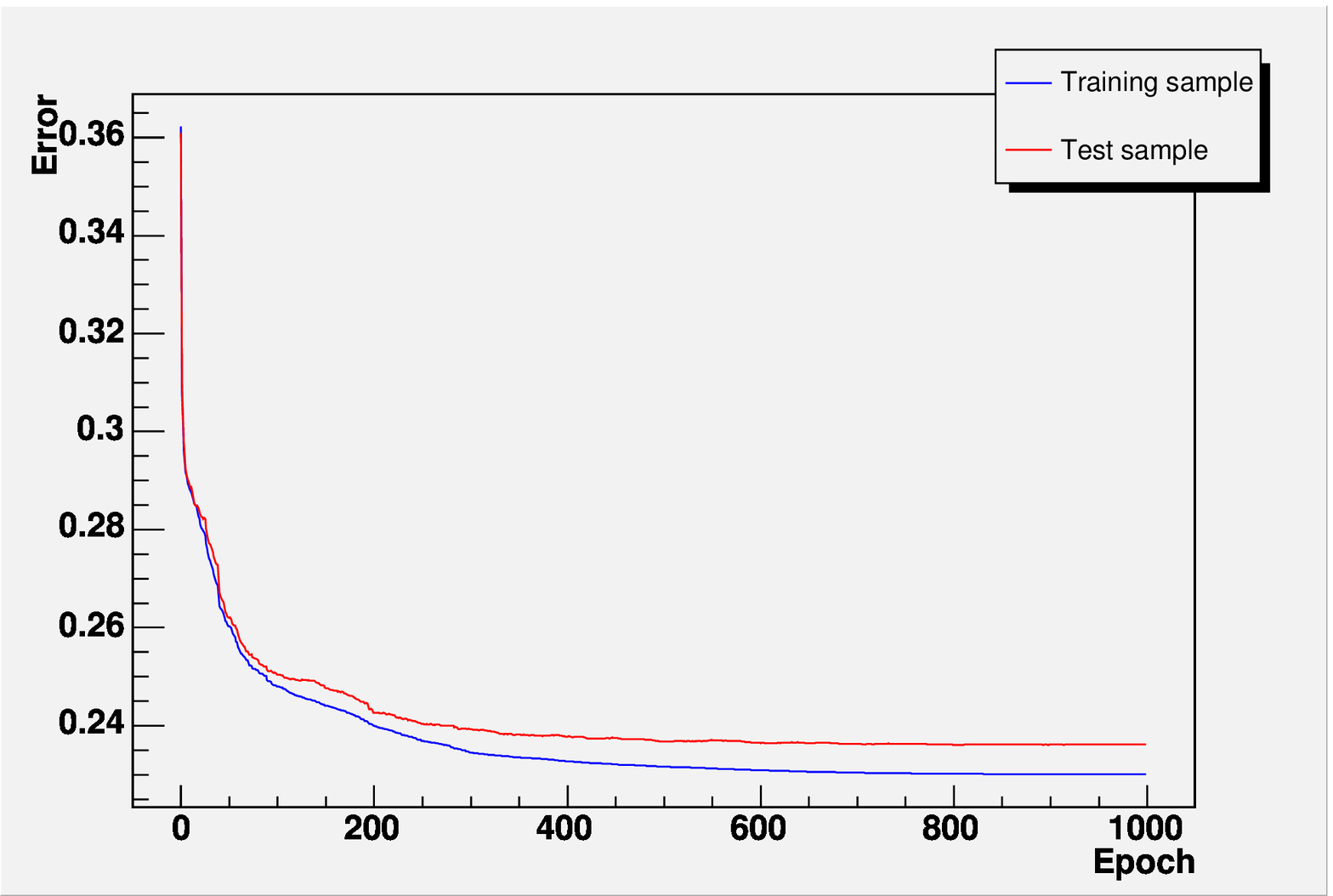}}}
  {\subfigure[The histogram of distributions for gamma and hadron parameters]{\includegraphics[width=6cm]{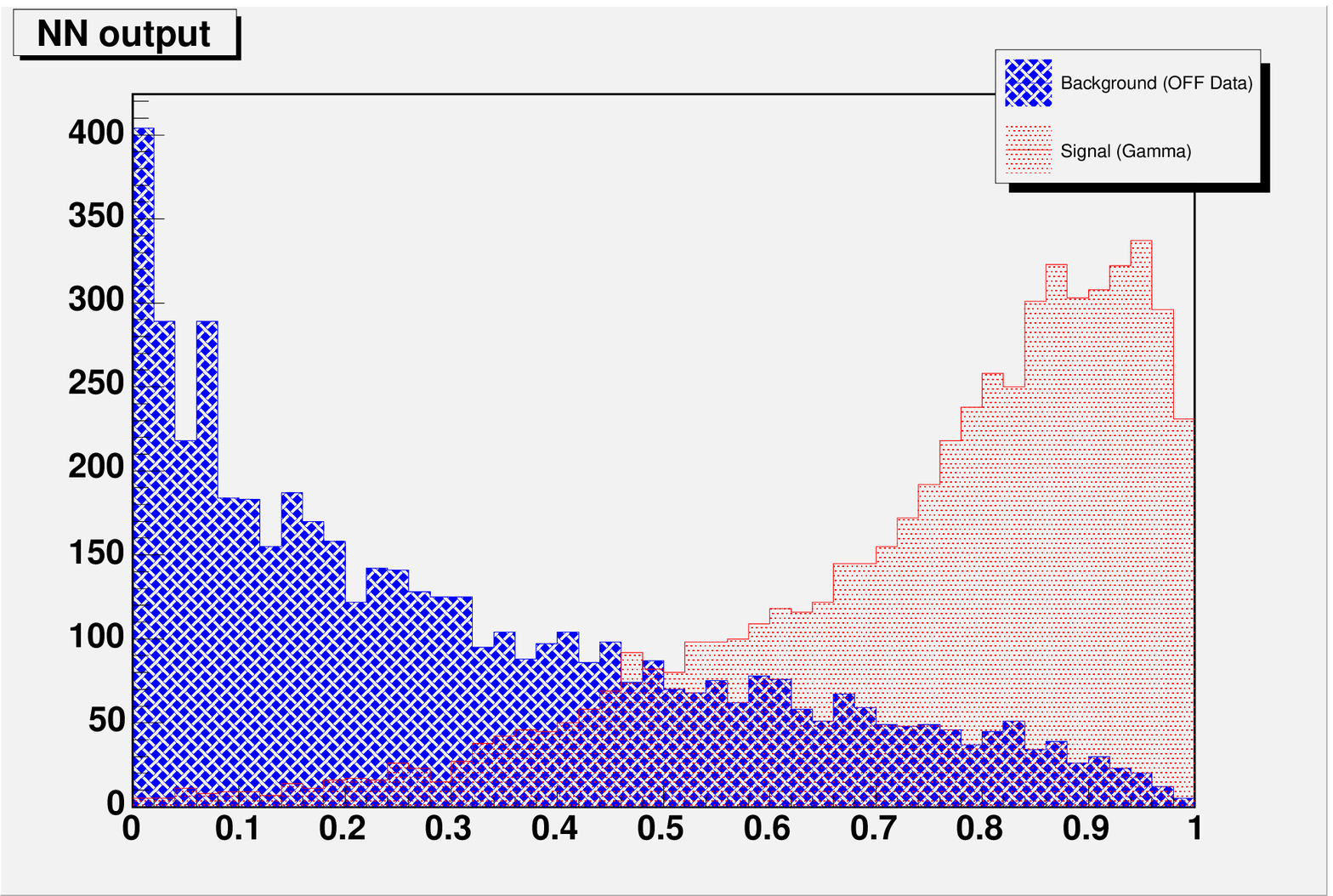}}}
  \caption{MLP\label{mlp} classification results using the BFGS default learning method.}
\end{figure}
Weights are put randomly at the beginning of the training session and then adjusted from the following runs in order to minimize errors (back-propagation). Errors at cycle $i$ are defined as: $err_i = \frac{1}{2} \; o_i^2$ where $o_i$ is the error of the output node.
Data to input and output nodes are transferred linearly, while for hidden layers they use a sigmoid (usually: $\sigma(x) = 1/(1 + \exp(-x))$).

We have tested the same network using different learning methods proposed by the code authors, as for example the so called "Stochastic minimization", based on the Robbins-Monro stochastic approximation, but the default "Broyden, Fletcher, Goldfarb, Shanno" (BFGS) method has proved to be the quickest and with the best error approximation.

Figures \ref{mlp}.a and \ref{mlp}.b represent a possible output when using the ROOT package on those data. The first one depicts the error function for each run of the network, comparing the training and the test data. Note that the greater is the number of runs, the better the network behaves.
The second one shows the distributions of output nodes, that is how many times the network decides to give a value near to "0" or to "1".

\section{Self-Organizing Tree Algorithm (SOTA)}
The Self-Organizing Tree Algorithm \protect\cite{sota} is an unsupervised neural network which implements a growing hierarchical clustering and is based on the self organising map network \protect\cite{p2}.
It hierarchically clusters the data into a binary tree of natural groups that exist in the data set.
Initially the tree consists of one root node linked to 2 child cells. All the input events are randomly distributed between the 2 initial child nodes. The tree grows by expanding the child node having the most heterogeneous population of associated inputs. Two new descendants are generated from this heterogeneous cell that changes its state from cell to node. The tree then grows by descending the cells into child nodes until each cell has one single input sequence, producing a complete classification of the sequences.%
\begin{figure}[hb]
\centering
\includegraphics[width=8cm]{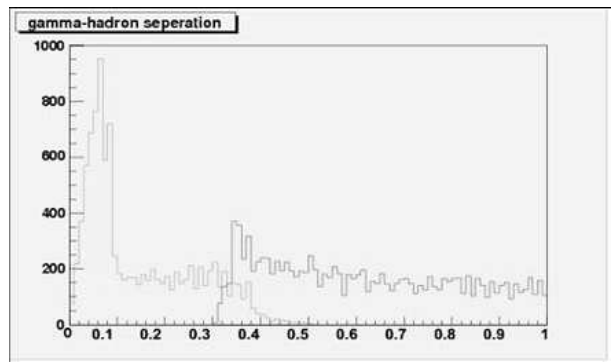}
\caption{SOTA\label{sotares} classification results.}
\end{figure} 
Alternatively, the expansion can be stopped at the desired level of heterogeneity in the cells, producing in this way a classification of sequences at a higher taxonomic level.

This kind of classification could be useful for astrophysics when a multi-event separation is needed on the same dataset, that is when multiple particles have been detected simultaneously and the analysis software should assign them a label (as "proton", "muon", "gamma", etc.).
They are also used as a data mining tool to explore the natural groups that exists in data sets.
Using the MAGIC datasets the tree has grown up to 10 levels, with the training sets taken from the MonteCarlo simulations (Figure \ref{sotares}).
This approach gives a hierarchical view of data, is robust for noisy data and is faster than traditional hierarchical clustering.

\begin{figure}[t]
\centering
\includegraphics[width=9cm]{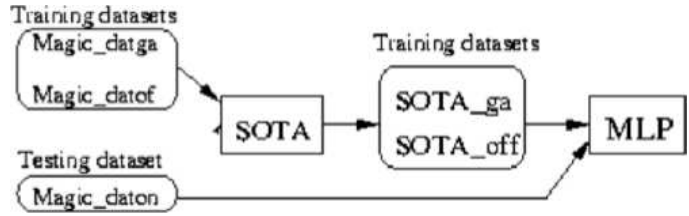}
\caption{Data\label{schema1} flow for a SOTA-MLP network using MonteCarlo datasets.}
\end{figure}

\subsection{SOTA + MLP}
Figure \ref{schema1} shows the combination of SOTA with MLP for the separation task.
The SOTA method is applied to the initial MonteCarlo datasets (gamma, ON and OFF) to find the natural clusters that exist in the datasets.
The SOTA tree produced two clusters of gamma and hadron which are used to train the MLP.%
\begin{figure}[ht]
\centering
\includegraphics[width=8cm]{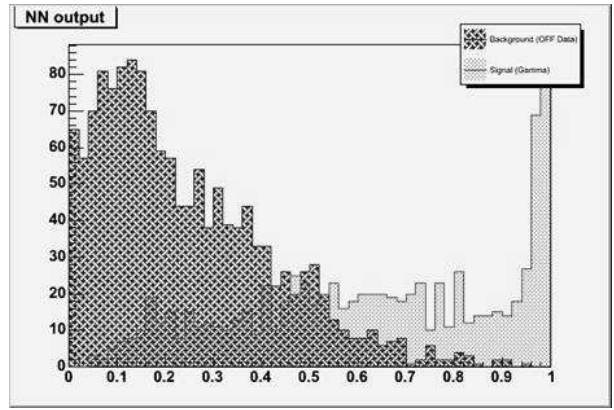}
\caption{A\label{sotamlp} preliminary result using an MLP feeded with SOTA labelled datasets.}
\end{figure}
The SOTA cluster emulates the data distribution of the patterns, thus reducing the number of events in the training set. The use of these clustered data can result in fast training for the MLP. The trained MLP network is then used to perform testing through the ON dataset and producing hadron probability for each event.

The preliminary results for this approach are shown on Figure \ref{sotamlp} where we can notice a better separation in the histograms respect to the non-treated MLP results (Figure \ref{mlp}.b).

\section{Discussion}
In this article we classified the gamma ray data using MLP and SOTA. Both MLP and SOTA shown some good classification results. The algorithms used here suggest that a complex problem could not be solved using standalone methods even if they are suitable for a large part of other data analysis problems.

SOTA algorithm clusters the data set into groups thus reducing the number of events in the training set. This can be useful for the MAGIC experiment where there are overwhelming events to be classified. MLP based on supervised technique identifies the group labels, but the training session could be longer. By combining SOTA with MLP we can  significantly decrease the training period and yield better classification results.

The  work can be further extended by using combination of different models in both self-organizing networks and supervised networks. Future experiments can be done using Growing Cell Structures and Growing Neural Gas models \protect\cite{gng} in the unsupervised category. In supervised networks, tests can be performed by using probabilistic networks and MLP trained with fast back-propagation.

\end{document}